\documentclass{llncs}
\usepackage{makeidx}  
\usepackage{epsfig}
\usepackage{amsmath}
\usepackage{graphicx}
\input{psfig.sty}
\input{epsf.sty}
\input mssymb.tex

\def\R{\Bbb R}

\begin{document}
%
%
\pagestyle{empty}
\addtocmark{Guarding a Set of Line Segments} 
%
\mainmatter              
\title{Patrolling a Street Network is Strongly NP-Complete
but in P for Tree Structures}
\titlerunning{Guarding a Set of Line Segments}
%

\author{Valentin E. Brimkov}

\authorrunning{Valentin E. Brimkov}

\institute{Mathematics Department, SUNY Buffalo State College, Buffalo, 
NY 14222, USA\\
\email{brimkove@buffalostate.edu}
}

\maketitle              

\begin{abstract}
We consider the following problem:
Given a finite set of straight line segments in the plane, determine the positions of a minimal number
of points on the segments, from which guards can see all segments.
This problem can be interpreted as looking for a minimal number of locations of policemen, guards, cameras or other sensors, that can observe a network of streets, corridors, tunnels, tubes, etc.  
We show that the problem is strongly NP-complete even for a set of segments with a cubic graph structure,
but in P for tree structures.

\smallskip

\end{abstract}
 
\section{Introduction}
\label{intro}

As an earliest source related to art-gallery problems authors usually refer to a question 
posed by Victor Klee at a conference in  1973: 
{\em How many guards are needed to patrol an art gallery with $n$ walls?} 
Actually, related studies have started much earlier by introducing the 
concepts of starshapedness and visibility (Brunn 1913 \cite{brunn}).
One of the first results is a Helly-type theorem of Krasnosel\'skii from 1946 
(Krasnosel\'skii's art gallery theorem~\cite{krasnoselski}),
that characterizes starshaped compact sets in $\R^n$.

Soon after Klee asked his question, in 1975, Chv\'atal proved that $\lfloor \frac{n}{3} \rfloor$ 
guards are sufficient to guard any polygon \cite{chvatal}.
Shorter proof was provided in \cite{fisk}.
Since then and especially in recent decades, art-gallery problems attract an increasing interest. 
Structural, algorithmic, and complexity results have been obtained for a great variety of art-gallery problems.
For getting acquainted with many of these the reader ir referred to the monograph of Joseph O'Rourke \cite{orourke}
and the more recent one of Jorge Urrutia \cite{urrutia}.  

Several works are devoted to guarding the facets of a planar praph (see, e.g., \cite{bkl,kz}  and the bibliography therein).
Others consider the problem for finding a minimal number of guard locations  in the plane that can observe 
a set of line segments (see, e.g., \cite{urrutia}). 
In the present paper we  consider a variant of an art-gallery problem which can informally be stated as follows.

\smallskip

\noindent Guarding a Set of Segments (GSS)

\noindent {\em Given a finite set of straight line segments in the plane, determine the positions of a minimal number
of points on the segments, from which guards can see all segments.} 

\smallskip

This problem can be interpreted as looking for a minimal number of locations of policemen, guards, cameras or other sensors, that can observe a network of streets, corridors, tunnels, tubes, etc.  

\begin{figure}[t]
\begin{center}
\begin{tabular}{c c c}
\includegraphics[width=4.5cm]{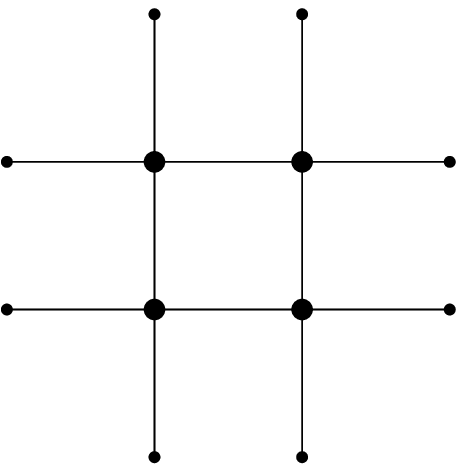} & \hspace{10mm} & \includegraphics[width=4.5cm]{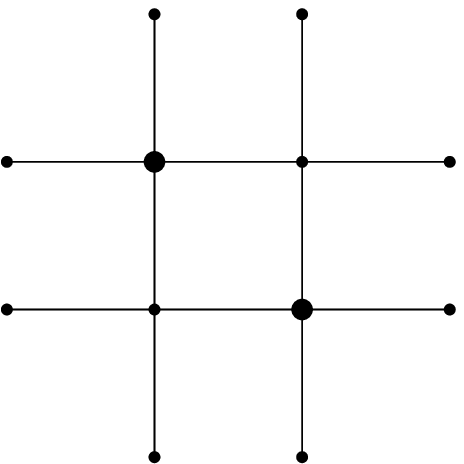} 
\end{tabular}
\end{center}
\caption{{\em Left:} Any minimal vertex cover of the given plane graph requires four vertices. One of them is 
marked by thick dots.  
{\em Right:} Two vertices can guard the same graph. One optimal solution is exhibited.}
\label{f1}
\end{figure}

This problem is germane to the set cover (SC) and vertex cover (VC) problems, 
which are fundamental combinatorial problems playing an important role in complexity theory.  
GSS can be formulated as a special case of the set cover problem (see Section~\ref{preliminary})
and, under certain conditions, as a vertex cover problem, as well.  
However, in general, GSS and VC are different, as Figure~\ref{f1} demonstrates. 
It is well-known that both SC and VC are NP-complete. 
Thus, it is interesting to know if GSS, being a particular case of SC and  similar to VC,
is NP-complete or not. 

In the present paper we obtain two main results. 
First, we show that GSS is strongly NP-complete, even if the graph corresponding to the set of segments is a cubic graph.
We also design an $O( p \log p)$-time algorithm that finds an optimal solution for the case when the set of segments features a tree structure (here $p$ is the number of segment intersections).

The paper is organized as follows. In the next section we introduce some notions and denotations, recall  some useful facts, 
and provide a formal statement of the problem.  
In Section~\ref{procedures} we describe certain useful data structures and procedures to be used in the sequel.
In Section~\ref{NP} we prove that GSS is strongly NP-complete.  
In Section~\ref{tree} we devise an  $O( p \log p)$-time algorithm that finds an optimal solution when
the graph corresponding to the set of segments is  a tree. 
Another polynomially solvable subclass of GSS is considered, as well. 
We conclude with some final remarks in Section~\ref{concl}.

\section{Preliminaries}
\label{preliminary}

\subsection{Basic Definitions and Facts}

In this section we fix some denotations to be used throughout the paper and recall
some notions and well-known results for further use.

\smallskip

For a set $A \subseteq \R^2$, by $|A|$ we denote its {\em cardinality} and by $d(A)$ 
its {\em diameter} defined as $d(A)=\max_{x,y \in A}||x-y||$,
where $||.||$ is the Euclidean norm. 
For $x,y \in \R^n$, $\rho(x,y) = ||x-y||$ is the Euclidean distance between $x$ and $y$.
Given two sets $A,B \subseteq \R^n$, $\rho(A,B) = \inf_{x,y} \rho(x,y), x \in A, y \in B \}$ is the Euclidean distance between them.
A straight line segment with end-points $X$ and $Y$ will be denoted by $\overline{XY}$,
and its length by $|XY|$.

A compact set $M \subset \R^2$ is called {\em star-shaped} (or {\em star}, for short)
if there is at least one point $c \in M$, called {\em star center}, 
such that for any point $x \in M$, the segment $\overline{cx} \subseteq M$. 

A graph is called a {\em star-graph} if all its vertices, possibly except one (the center of the star-graph), 
have degree one.

The problem we consider is closely related to two fundamental combinatorial problems: the set cover problem and the vertex cover problem.

Given a finite set $U$ and a family $F$ of subsets of $U$, a {\em cover} of $U$ is a subfamily $C \subseteq F$ whose union is $U$. 
The set cover problem (SC) has as an input $U$  and $F$ defined above, and  a positive integer $K \leq |F|$.
The question in the decision version of the problem is whether there is a cover $C$ with $|C| \leq K$.
In the optimization version one looks for a cover with a minimal number of elements.
 
Given a graph $G$, a {\em vertex cover} of $G$ is a set $C$ of vertices of $G$, such that every edge of $G$ is incident
to at least one vertex of $G$. 
The vertex cover problem (VC) has as an input a graph $G=(V,E)$ and a positive integer $K \leq |V|$.
The question in the decision vertex cover problem is whether there is a vertex cover $C$ with $|C| \leq K$. 
In the optimization version one looks for a vertex cover with a minimal number of elements.
It is well-known that the decision/optimization SC and VC are NP-complete/hard \cite{karp72} (see also \cite{GJ}). 

A F\'ary embedding of a planar graph in the plane is an embedding in which all edges are straight line segments.
It is well-known that every planar graph admits a F\'ary embedding.
In the NP-completeness proof in Section~\ref{NP} we will use the following well-known result of
Fraysseix, Pach, and Pollack.
\begin{lemma} \cite{FPP}
Given a planar graph on $n$ vertices, there is an $O(n \log n)$ time $O(n)$ space algorithm
that computes a F\'ary embedding of $G$ on the $(2n-2) \times (n-2)$-integer grid.  
\label{FPP}
\end{lemma}

Finally, we recall that complexity theory distinguishes between problems 
with and without numeric data.  
For problems of the latter type the largest number appearing in the input can be bounded by a polynomial in the 
problem size, while for problems of the former type this is not possible. 
Such problems are sometimes called {\em number problems}. 
Thus, the set cover and vertex cover problems are not number problems,
while GSS is, since, in general, the coordinates of a segment endpoint is in no way be bounded 
by a polynomial in the number of segments.     

It is a well-known fact of early complexity theory that the hardness of some number problems is due to the possible  presence of large 
numbers in the input rather than to their combinatorial structure.  
Some number problems are polynomially solvable if the largest number in their input is bounded by the problem size,
others remain NP-complete/hard even under such a condition. 
Problems of the latter type are known as {\em strongly NP-complete/hard}.  
Clearly, non-number NP-complete/hard problems are strongly NP-complete/hard.
Thus, both the set cover and the vertex cover problems are strongly NP-complete.
In Section~\ref{NP} we will show that GSS, although being a number problem, is strongly NP-complete, as well.  
The authors suppose that every possible reader of this paper would be well-familiar with the 
basics of the theory of NP-completeness, including the above notions which we recalled only for
the sake of completeness. 
Formal definitions and any details are available in \cite{GJ} (see also \cite{cormen}).

\subsection{Further Notations and Problem Statement}  

Let $S=\{s_1,s_2,\dots,s_n\}$ be a set of segments in  the plane. 
Denote by $\bar S = \cup_{s \in S} s$ the set of all points of segments in $S$.
Let $I=\{ v_1, v_2,\dots,v_m \}$ be the set of all intersection points and $J$ the set of all end points 
of segments of $S$. Denote $V=I \cup J$. $V$ will be called {\em the vertices} of $\bar S$. 
For technical simplicity only and without loss of generality we will assume that if two collinear segments
intersect, they have only one intersection point (their common endpoint) which also belongs to at least one more segment
(otherwise, it is trivial to discover and merge into one any two adjacent collinear edges that are not affected by another edge at their intersection). 

Let $G_{\bar S}=(V,E)$ be the plane graph whose embedding is $\bar S$.

For a segment $s \in S$, let $E_s$ be the set of edges of $G_{\bar S}$ contained in $s$.

The sets of edges incident to a vertex $u$ in $G_{\bar S}$ 
will be denoted by $E_u$.
\begin{figure}[t]
\begin{center}
\includegraphics[width=6.9cm]{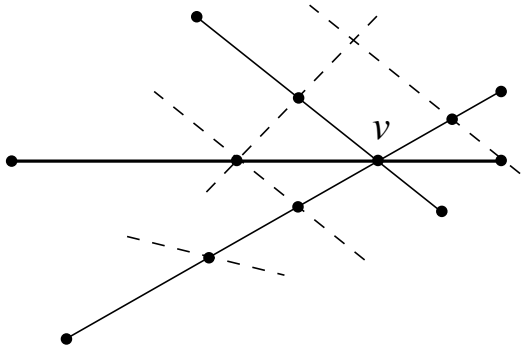}
\end{center}
\caption{A pseudo-star graph centered at vertex $v$.}
  \label{f2a}
\end{figure}

For $v \in V$, let $S_v$ be the set of segments from $S$ containing $v$. 
Denote $\bar S_v = \cup_{s \in S_v} s$. 
Clearly, the set $\bar S_v$ is star-shaped.
Its corresponding subgraph of $G_{\bar S}$ 
is a tree  whose vertices (possibly but one) have degree 1 or 2. 
We will call such graphs {\em pseudo-star graphs} (see Figure~\ref{f2a})
and denote their set of edges by $F_u$.

Being a center of a star $\bar S_v$, $v$ is connected to every point of $\bar S_v$ by a segment contained in
$\bar S_v$. Therefore, one can say that $v$ ``sees" every point of $\bar S_v$.

With the above preparation, the problem of guarding a set of segments can be formulated as follows.

\smallskip

\noindent {\bf Guarding a Set of Segments (GSS)}

\noindent {\em Fing a minimal  (by number of elements) subset of vertices $\Gamma \subseteq V$,    
such that $\cup_{v \in \Gamma} S_v = \bar S$. }

\smallskip

In other words, one has to locate a minimal number of guards at the vertices of $\bar S$,
so that every point of $\bar S$ is seen by at least one guard.   

\begin{remark}
It is easy to see that the requirement to locate guards at vertices is not a restriction of the generality:
every non-vertex point on a segment $s$ can see the points of $s$ only, while each of the vertices on $s$ 
can see $s$ and possibly other segments.   
Thus, looking for a minimal set of guard locations is equivalent to finding a minimal 
number of stars (centered at vertices of $S$) whose union is $\bar S$.
\label{rr}
\end{remark}
In view of the above remark,
GSS admits formulation in terms of a set-cover problem, as follows.

\smallskip

\noindent {\bf Set Cover Formulation of GSS}

{\em Let $S$ be a set of segments and $V$ the set of their end-points and intersections.
Find a minimal subset of  vertices $\Gamma \subseteq V$,    
such that $\cup_{u \in \Gamma} S_u = S$. }

\smallskip

Note that, given a set of segments $S$ as a problem input, the set $V$ is not readily available.
However, it is efficiently computable in $O(m + n \log^2 n/\log\log n)$ time (see Procedure~(A) 
in Section~\ref{procedures}). 

\subsection{How Many Guards Are Always Sufficient to Guard a Set of Segments?}
As already mentioned, the answer to Klee's question about guarding a polygon was first given by Chvatal \cite{chvatal}. 
Regarding GSS, clearly a single segment requires one guard. 
For a set of more than one segment, an answer to Klee's question is given by the following proposition.

\begin{figure}[t]
\begin{center}
\includegraphics[width=9.6cm]{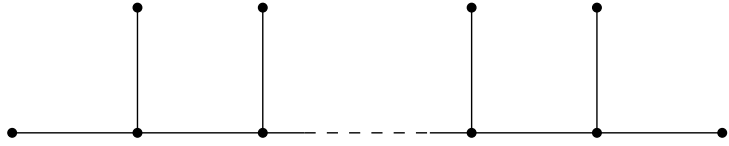} 
\end{center}
\caption{Illustration to the second part of Proposition~\ref{prop1}.}
  \label{f2b}
\end{figure}

\begin{proposition}
$n-1$ guards are always sufficient to guard $n>1$ segments in the plane.
This number of guards is the best possible for some subclasses of GSS.
\label{prop1}
\end{proposition}
{\bf Proof} \
For any $S$ with $|S| \geq 2$, there must be a vertex $v$ of $\bar S$ that is an intersection of at least two segments.
Then a guard placed at $v$ will patrol all these segments. 
Placing a guard at one of the endpoints for all other segments provides a guarding set with no more than $n-1$ guards.
This number is the best possible for the set of segments in Figure~\ref{f2b}.
\qed

\section{Some Useful Data Structures and Procedures} 
\label{procedures}

In this section we present some useful data structures and procedures to be used in the following sections.

\smallskip

\noindent {\bf Procedure~(A)} ($I$ computation)

\noindent {\em Input:} Set of segments $S=\{s_1,s_2,\dots,s_n\}$.

\noindent {\em Output:} Set $I=\{ v_1, v_2,\dots,v_m \}$ of all intersection points of segments from $S$.

\smallskip

Finding $I$ is a fundamental and extensively studied problem in computational geometry.
Well-known are an $O((n+m) \log n)$-algorithm of Bentley and Ottmann \cite{bentleyOttmann} 
and the more efficient $O(m + n \log^2 n/\log\log n)$-algorithm of Chazelle \cite{chazelle}.
See also Ch. 7.2 of \cite{PreparataShamos}.
These algorithms also provide the sets $S_u$ of segments intersecting at $u$. 

\smallskip

\noindent {\bf Procedure~(B)} ($E_s$ computation)

\noindent {\em Input:} Set of segments $S=\{s_1,s_2,\dots,s_n\}$.

\noindent {\em Output:} Ordered set of edges of $\bar G_S$ on every $s \in S$.

\smallskip

The procedure consists of the following steps.

\begin{enumerate}
\item[1.]
Using Procedure~(A), compute the set $I=\{ v_1, v_2,\dots,v_m \}$ 
and lists $S_u$ for all $u \in V$.  
Let $V = I \cup J =\{ v_1, v_2,\dots,v_m,v_{m+1},\dots,v_p \}$ 
\item[2.]
Initialize $n$ double-indexed lists $L_1,L_2,\dots,L_n$ to empty. 
\item[3.]
Consecutively read $S_{v_1}, S_{v_2},\dots,S_{v_p}$.
Add an element $l(i,j)$ to a list $L_k$ as soon as a vertex $v_j$ is met in a list $S_u$.  
The first index $i$ indicates the consecutive number of $l(i,j)$ in $L_k$.
\item[4.]
After the last element of $S_{v_p}$ is processed, sort all lists $L_i$.
\item[5.]
From the obtained sorted lists, for every $s$ reconstruct $E_s$ using the indexes $j$.  
\end{enumerate}

\begin{proposition}
Procedure~(B) computes all sets $E_s$ in $O(p \log p)$ time. 
\label{prop2}
\end{proposition}
{\bf Proof} \
The correctness of the procedure is self-justified, so only its time-complexity needs explanation.

Step 1 can be performed in $O(m + n \log^2 n/\log\log n)$ time and Step 2 takes $O(n)$ time. 

We now evaluate Step 3. 
For every $s \in S$ we obviously have $|S_v| \leq |E_v|$. 
Hence,
\begin{equation}
\sum_v |S_v| \leq \sum_v |E_v|
\label{e1}
\end{equation}
Since each edge $e=(u,v) \in E$ is incident to exactly two vertices $u$ and $v$, 
we have that $e$ is contained in exactly two sets $E_u$ and $E_v$. 
Then $\sum_v |E_v| = 2|E|$. Since $\bar G_S$ is planar, $|E|=O(|V|)=O(p)$. 
Then from (\ref{e1}) we obtain
\begin{equation}
\sum_v |S_v| \leq 2|E| = O(p).
\label{e2}
\end{equation}
Thus, Step 3 takes $O(p)$ time.

In Step 4, sorting a list $L_i$ takes $O( |L_i| \log |L_i| )$ time. 
Then, keeping in mind (\ref{e2}), the overall  time complexity is
$$
O( \sum_{i=1}^n |L_i| \log |L_i| ) = O( (\sum_{i=1}^n |L_i|) \log p ) = O( p \log p ).
$$   
Step 5 consists of relabeling of the edges in the lists $L_i$ and takes $O(p)$ time.
\qed

\smallskip

\noindent {\bf Procedure~(C)} ($F_u$ computation)

\noindent {\em Input:} Set of segments $S=\{s_1,s_2,\dots,s_n\}$.

\noindent {\em Output:} For every vertex $u \in V$, compute the set of edges/vertices visible from $u$.

\smallskip

The above procedure is directly implied by Procedure~(B).

\smallskip

\noindent {\bf Procedure~(D)} ($E_u$ computation)

\noindent {\em Input:} Set of segments $S=\{s_1,s_2,\dots,s_n\}$.

\noindent {\em Output:} For every vertex $u \in V$, compute the set of edges incident to $u$.

\smallskip

By Procedure~(A) we find  all vertices $u$ and the corresponding sets $S_u$.
For each $s \in S_u$, by Procedures (B) and (C) we can find the list of vertices/edges  on $s$ and identify the position
of $u$ in this list in $O(p \log p)$ time.
Then the neighbors of $u$ will provide the edges of $E_u$.

\section{Guarding a Set of Line Segments is Strongly NP-Complete} 
\label{NP}

In this section we prove the following theorem.

\begin{theorem}
The GSS problem is strongly NP-complete.
\label{th1}
\end{theorem}
{\bf Proof} \
We will consider GSS in its Set Cover form:

\smallskip

\noindent {\em Guarding Set of Segments  (GSS):} 

\noindent {\em Instance:} \ A set of segments $S=\{s_1,s_2,\dots,s_n\}$ together with the set of 
their endpoints and intersections $W = \{ v_1, v_2,\dots,v_p \}$, and a positive integer $K \leq p$. 

\noindent {\em Question:} \ Is there a set $\Gamma \subseteq W$ with $|\Gamma| \leq K$, 
such that  $\cup_{u \in \Gamma} S_u = S$?

\smallskip

It is trivial to show that ${\rm GSS} \in {\rm NP}$: Given a candidate solution $\Gamma$,
one can check in polynomial time if $|\Gamma| \leq K$ and if each segment in $S$ contains an element of $\Gamma$. 

In the rest of this section we exhibit a polynomial reduction to GSS of the following problem known to be strongly
NP-complete \cite{GJ2}.

\smallskip

\noindent {\em Vertex Cover in a Planar Cubic Graph (3PVC)}

\noindent {\em Instance:} \ A planar cubic graph $G=(V,E)$ 
(i.e., a planar graph whose vertices have degree no greater than 3) and a positive integer $M \leq |V|$.

\noindent {\em Question:} \ Is there a {\em vertex cover} for $G$ of size no greater than $M$, i.e., 
a subset $W \subseteq V$ with $|W| \leq M$, and such that for every edge $(u,v) \in E$, 
at least one of $u$ and $v$ belongs to $W$? 

\smallskip

The idea of our polynomial reduction is as follows. 

First we obtain in $O( n \log n )$ time and $O(n)$ space a Fa\'ry embedding of $G$ onto a 
$(2n-4) \times (n-2)$-grid in the plane,
using the algorithm of Fraysseix, Pach, and Pollack (Lemma~\ref{FPP}). 
However, the obtained embedding $G'$ may have collinear vertices. 
Therefore, as an essential part of our construction, we deform in polynomial time and space $G'$ to a plane graph $G''$, which is essentially the same as $G'$ but features no collinearities. 
Moreover, the size of the coordinates of the deformed vertices is polynomial in $n$.
On the so-constructed graph $G''$, the GSS problem turns out to be equivalent to the original 3PVC problem with the same bound $M$,
which implies the strong MP-completeness of GSS.
More detailed description is given next.

\subsection*{Construction of GSS Instance}

As already mentioned, the first step of the reduction is 
embedding $G$ in the plane using the algorithm of Lemma~\ref{FPP}.
Let $G'=(V',E')$ be the obtained embedding.
W.l.o.g., assume that the embedding is in the $(2n-4) \times (n-2)$-grid whose lower-left corner is at the origon of the coordinate system.

In order to destroy all possible vertex collinearities in $G'$, first we need to identify them.
For this, we need some data organization and processing.   

\subsubsection*{Finding collinearities.}

We have $n$ edges determining straight lines. 
Put every line  in the form $y = m x + b$, where the slope $m$ and the intercept $b$ are rational numbers in their lowest terms (i.e., irreducible fractions). 
This requires $O(n \log n)$ operations overall. (The factor $\log n$ comes from putting $m$ and $b$ in 
lowest terms which requires finding the $gcd$ of the numerator and the denominator by the Euclidean algorithm.)  
Since every line is determined by two points with coordinates not exceeding $2n-4$, the size of $m$ and $b$ 
is polynomial (linear) in $n$. 
If a line is vertical or horizontal, its equation is $x=a$, resp. $y=b$, where $a$ and $b$ are integers not exceeding
$2n-4$ and $n-2$, respectively. 

To each line an index is associated, indicating to which edge it corresponds. 
Note that the same line may correspond to different edges. 

Now sort the list of $m$'s ($O(n \log n)$ operations). 
This puts in the sorted list together those lines having the same slope. 
Then sort one more time with respect to $b$-parameter each group of identical $m$'s.  
This takes $O(n \log n)$ operations overall. 
To see this, let $k_1,k_2,\dots,k_m$ be the numbers of lines in the different groups with respect to the line slopes. 
We have $k_1+\dots+k_m=O(n)$. Then the overall time of the $m$ sorting procedures is 
$$
O(k_1 \log k_1 + k_2 \log k_2 + \dots + k_m \log k_m) = 
$$
$$
O(k_1 \log n + k_2 \log n \dots +  k_m \log n) = 
$$
$$
O(k_1 + \dots + k_m) \log n = O(n \log n). 
$$
Since the sortings by $m$ and $b$ are performed consecutively, 
their running times are just added in the overall running time evaluation. 
Clearly, the edges corresponding to equivalent pairs $(m,b)$ will lie on the same straight line.
The more trivial case of vertical or horizontal lines is handled analogously.
 
Next, sort the discovered collinear edges in each group by the left endpoint. 
This requires $O(k \log k)$ operations where $k$ is the number of edges in a group of collinear edges.
The overall time complexity of this step is $O(n \log n)$ as well, by the argument used above to 
evaluate the time complexity of sorting groups by the parameter $b$. 
\begin{figure}[t]
\begin{center}
\includegraphics[width=7.5cm]{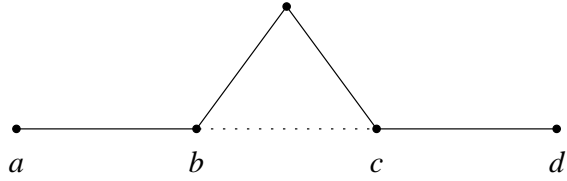} 
\end{center}
\caption{The edges $(a,b)$ and $(c,d)$ belong to the same straight line but do not feature  collinearity in
the considered sense.}
  \label{f3}
\end{figure}

Finally, for every ordered set $L$ of collinear edges, 
we group together those that exhibit a connected sequence of edges over the line.
(Note that, for example, two edges $(a,b)$ and $(c,d)$ may belong to the same straight line but there may be 
no edge or a sequence of edges on the same line that connect the vertices $b$ and $c$, see Fifure~\ref{f3}). 
For this it suffices to check for every two consecutive elements $(a,b)$ and $(c,d)$ of $L$ whether 
the vertices $b$ and $c$ are identical. So, the last step requires $O(n)$ operations.

Thus we have computed in $O(n \log n)$ time all sets of consecutive collinear edges, 
sorted as they appear on a line.

\subsubsection*{Moving vertices.}

Having all colliniarities discovered, we get rid of them as follows.

\begin{figure}[t]
\begin{center}
\begin{tabular}{c c c}
\includegraphics[width=6.5cm]{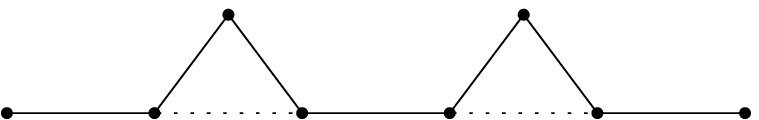} & \hspace{6mm} & \includegraphics[width=4.5cm]{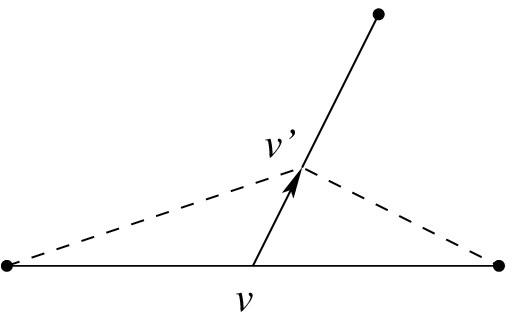} 
\end{tabular}
\end{center}
\caption{Illustration to Rules 1, 2, and 4 for moving vertices.}
  \label{f4}
\end{figure}

Let $u_1,u_2,\dots,u_r$ be a sequence of collinear vertices of consecutive collinear edges
$(u_1,u_2),(u_2,u_3),\dots,(u_{r-1},u_r)$.
Let $l$ be the line to which the vertices and edges belong. 
Vertices are moved according to the following rules.
\begin{itemize}
\item[R1]
If $r$ is odd, we move every even numbered vertex, otherwise we do the same except for the last one.\\
All vertices to be moved are labeled. See Figures~\ref{f4}, left.
\item[R2]
All moved vertices are moved to the same half-plane with respect to the line $l$.  \\
See Figures~\ref{f4}, left.
\item[R3]
Each movement is at distance $\frac{1}{6n}$. 
\item[R4]
\begin{itemize}
\item[-]
A vertex of degree 2 is moved either horizontally or vertically.
If the line $l$ is vertical, then the move is horizontal; if $l$ is horizontal, then the move is vertical.
Otherwise, the vertical/horizontal option is chosen arbitrarily.  
\item[-]
A vertex of degree 3 is moved along the third edge that is noncollinear to the other two.\\
See Figures~\ref{f4}, right.
\end{itemize}
\end{itemize}

It is easy to see that by construction the size of the obtained graph $G''$ is linear in $n$ and 
is computed from $G'$ in time linear in $n$ (provided that all collinearities are found in the preprocessing phase).
What remains is to show that the vertex movements do not cause any new edge intersections and the obtained graph $G''$ 
has no collinearities (i.e., has no two collinear adjacent edges).  

\subsection*{Final Analysis}

We start by listing a useful technical fact that most probably belongs to the mathematical folklore.

\begin{lemma} (see, e.g., \cite{brimkov})
Let $S$ be a set of line segments in the plane.
Let $U = \{ v_1,v_2,\dots,v_m \}$ be the set of the segment endpoints and suppose that they are all integer.
Let $d(\bar S)$ be the diameter of $\bar S=\cup_{s \in S} s$. 
Clearly, $d(\bar S)= \max_{v_i,v_j} ||v_i-v_j||, 1 \leq i,j \leq m$.\\
For every $v_i \in U$, define $\eta_i(\bar S) = \min_{j,k \neq i} \rho(v_i,\overline{v_j v_k})$,
where $v_i \notin \overline{v_j v_k}$ and $\overline{v_j v_k}$ is the straight line through $v_j$ and $v_k$.
Let $\eta(\bar S) = \min_i \eta_i(\bar S)$. Then $\eta(\bar S) \geq 1/d(P)$.
\label{Ldiam}
\end{lemma}
Informally, the above lemma says that any integer endpoint is no closer to a segment or its line extension than
the reciprocal of the diameter of the endpoints.

We have that the diameter of the embedding $G'$ satisfies  
$$
d(\bar G) \leq \sqrt{(2n-4)^2 + (n-2)^2} < \sqrt{(2n)^2+n^2} = \sqrt{5}n < 3n.
$$
Then by Lemma~\ref{Ldiam}, $\eta(\bar S) > \frac{1}{3n}$.

Now, with a reference to the rules for vertex movement, we observe the following.  

First, note that all deformations of $G'$ are local and if a vertex is moved, it is moved only once. 
Let $u$ be a vertex to be moved. 
Let $l$ be a straight line determined by arbitrary two other vertices of $G'$.
By Lemma~\ref{Ldiam},  the distance from $u$ to $l$ is strictly greater than $\frac{1}{3n}$.
Therefore, when $u$ is moved at distance $\frac{1}{6n}$ to a new position $u'$,
the distance from $u'$ to $l$ will remain strictly  greater than $\frac{1}{6n}$.

By rule R4, when a vertex $u$ is moved, exactly two edges $(x,u)$ and $(u,y)$ adjacent to 
$u$ are moved together with it.
Before the move these are collinear, i.e., lie on a line $g$. 
According to Lemma~\ref{Ldiam}, any vertex not on $g$
is at distance from the line strictly greater than  $\frac{1}{3n}$.
Thus, after a move, all vertices not on $g$ will remain 
at distance greater than $\frac{1}{6n}$ from each of the segments  $(x,u')$ and $(u',y)$.
 
\begin{figure}[t]
\begin{center}
\begin{tabular}{c c c}
\includegraphics[width=4.5cm]{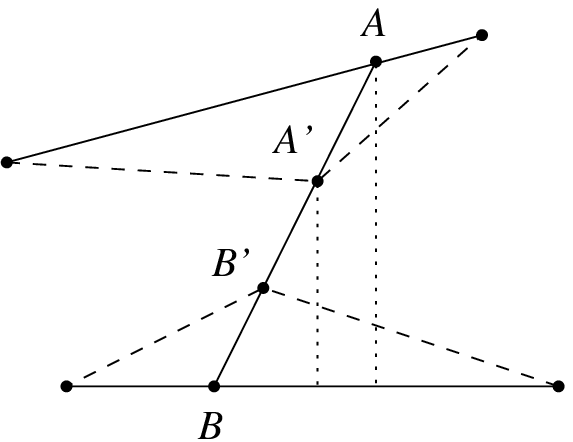} & \hspace{6mm} & \includegraphics[width=6.3cm]{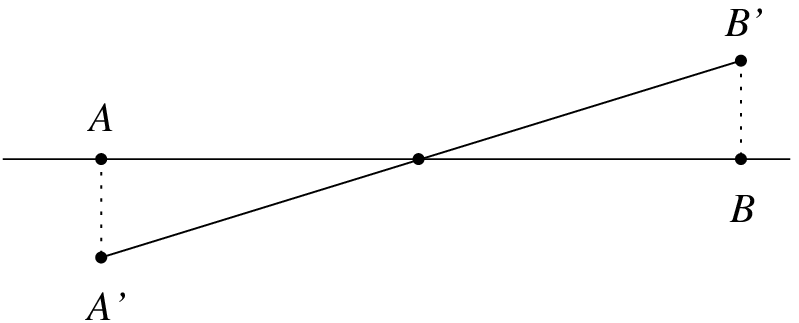} 
\end{tabular}
\end{center}
\caption{Illustrations to the proof of the Theorem~\ref{th1}.}
  \label{f5}
\end{figure}

Now let another vertex $v$ of $G'$ be moved at a later point. 
Similar reasoning as above makes clear that after the move $v$ remains at a distance greater than $0$ 
from both $(x,u')$ and $(u',y)$. 
Thus, when a vertex is moved, it does not cross or touch any edge of the current graph.
See Figure~\ref{f5}, left for illustration.

According to rules R1 and R2, after all labeled vertices are moved, all existing collinearities are removed.  
Moreover, no new collinearities can be introduced, as rule R2 excludes the scenario exhibited in Figure~\ref{f5}, right.   

The time complexity of the different steps of the reduction was  analyzed together with their description. 
Overall, it amounts to $O(n \log n)$. 

This completes the description of the polynomial reduction of 3PVC on a graph $G$ to a special instance of GSS
on the graph $G''$ with the same constant $M$.
By the construction of $G''$, 3PVC has a vertex cover of no more than $M$ vertices if and only if 
GSS admits a solution of no more than $M$ guards, which completes the proof. 
\qed

\smallskip

The proof of Theorem~\ref{th1} implies the following corollary.
\begin{corollary}
The GSS problem is strongly NP-complete for sets of segments with a cubic graph $\bar G$.  
\end{corollary}

\begin{figure}[t]
\begin{center}
\begin{tabular}{c c c}
\includegraphics[width=4.5cm]{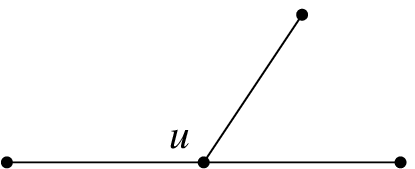} & \hspace{6mm} & \includegraphics[width=4.5cm]{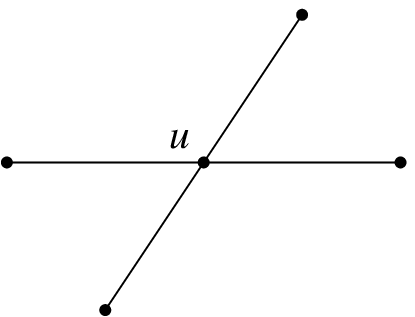} 
\end{tabular}
\end{center}
\caption{Illustration to Remark~\ref{r1}.}
\label{f6}
\end{figure}

We conclude this section with one more remark. 
\begin{remark}
Consider the class of GSS for which all vertices are of degree 2, 3, or 4, as the following conditions are met:
\begin{itemize}
\item[-]
If in the graph $\bar G$ a vertex $u$ has degree 3, two of the edges incident to $u$ are collinear. 
That is, $u$ is an intersection of two segments, as the intersection point is the endpoint of one of them. 
See Figure~\ref{f6}, left.
\item[-]
If $u$ has degree four, then two of the edges incident to $u$ are collinear, and the other two are collinear as well.  
That is, $u$ is an intersection of two segments, as the intersection point is internal for each of them.
See Figure~\ref{f6}, right.
\end{itemize}
If the above is the case, in the set cover formulation of GSS every family has at most two elements.
It is well-known that a set cover problem with this property can be solved in polynomial time \cite{GJ}.  
\label{r1}
\end{remark}

\section{Polynomial Algorithm for Guarding a Plane Tree} 
\label{tree}

As the general GSS problem is strongly NP-complete, we look for special subclasses of GSS for which a 
polynomial algorithm exists. 
A trivial but important observation is that if the number of intersections is comparatively small, the problem can be
solved efficiently.
More precisely, we have the following fact.
\begin{proposition}
Let the number of intersections $m=|I|$ satisfies $m = O(\log^c n)$, where $c$ is an arbitrary 
positive integer constant. Then GSS can be solved in time $O(n^{c+1})$.  
\label{prop3}
\end{proposition}
{\bf Proof} \
It is not hard to realize that any GSS instance admits an optimal solution
in which any guard location is an intersection of at least two segments (an intersection may belong to 
segment interior or may be its endpoint).  
By Procedure~(A) one can compute $I$ and $S_u$ for all $u \in I$ in $O(m + n \log^2 n/\log\log n)$ time. 
An exhaustive generation of all subsets of $I$ requires overall $O(2^m)= O(n^c)$ arithmetic operations. 
Once a subset $Q \subseteq I$ is generated, the union
$\cup_{u \in Q} S_u$ is computed and compared with $S$.
Both can be done in $O(n)$ time, which implies the result stated. 
\qed
 
\smallskip

Next we show that GSS can efficiently be solved if the $G_{\bar S}$ is a tree.   
For this, we need some preliminaries.

\subsection*{Appropriate Leaves}
\begin{figure}[t]
\begin{center}
\begin{tabular}{c c c}
\includegraphics[width=4.8cm]{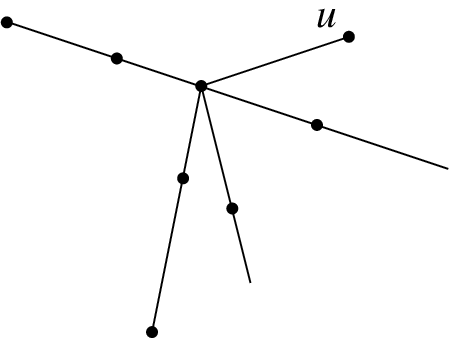} & \hspace{6mm} & \includegraphics[width=4.8cm]{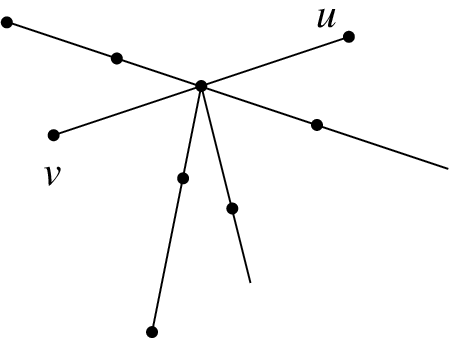} 
\end{tabular}
\end{center}
\caption{Illustration to the notion of appropriate leave. Leave $u$ is appropriate, and its counterpart $v$ in the right
figure is appropriate as well.}
\label{f7}
\end{figure}

Let $T = (V,E)$ be a tree.
For every vertex $u \in V$,  we can compute by Procedure~(D) in $O(|V| \log |V|)$ time 
the set $E_u$ of edges incident to  $u$.
If $|E_u|=1$, then $u$ is a leave of $T$.  
We call a leave $u$ {\em appropriate} if it is visible only from its parent or by another leave across the parent
(see Figure~\ref{f7}).  
For a given leave $u$, Procedure~(C) provides the set $F_u$ of all edges (and vertices) visible from $u$,
i.e., that see $u$. 
If $|F_u|=1$ or $2$, then $u$ is appropriate. 
So, all appropriate leaves of $T$ are computable in $O(|V| \log |V|)$ time.
We have the following lemma.
\begin{lemma}
Every tree $T$ representing a GSS problem has appropriate leaves. 
\label{L1}
\end{lemma}
{\bf Proof} \ 
We prove by induction on $|V|$. 
For a tree $T$ on two or three vertices the statement is obvious.
Assume that it is true for a tree on $k \geq 3$ vertices.
Remove an arbitrary leave $u$ together with the incident edge $(u,v)$ (without removing the parent $v$)
and consider the obtained tree $T'$.
It is a tree on $k$ vertices and by the inductional hypothesis has an appropriate leave $w$.
If $w \neq v$, we are done. 
Otherwise, let $p$ be the parent of $v$. 
Then $u$ is an appropriate leave in $T$, as $p$, $v$, and $u$ cannot be collinear,
therefore $v$ is the only vertex that sees $u$.   
\qed

\subsection*{Removing Edges}

Let $u \in V$ and $F_u$ be the set of edges visible from $u$. 
Let us remove from $E$ all edges of $F_u$  without their vertices.
This will turn $T$ to a forest on $|V|$ vertices, $|E|-|F_u|$ edges (set of edges denoted $E-F_u$), 
and $|F_u|+1$ components.  
We have the following lemma.
\begin{lemma}
Let $T=(V,E)$ be a tree representing a GSS problem and $u$ an appropriate leave of $T$.
Then the graph $(V,E-F_u)$ has either an empty set of edges of contains all appropriate leaves of 
$T$ except those adjacent to $u$.
\label{L2}
\end{lemma}
Follows from Lemma~\ref{L1} and the definition of an appropriate leave.

\smallskip

Finally, we list one more fact.    

\begin{figure}[t]
\begin{center}
\begin{tabular}{c c c}
\includegraphics[width=4cm]{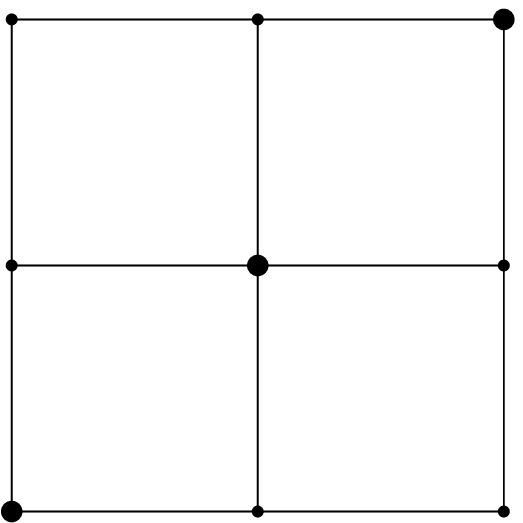} & \hspace{10mm} & \includegraphics[width=4cm]{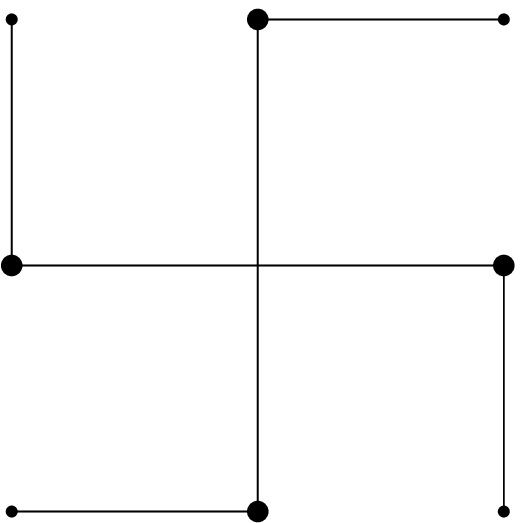} 
\end{tabular}
\end{center}
\caption{The graph on the left can be guarded by three verticess while its subgrapg on the 
right needs four guards.}
  \label{f8}
\end{figure}

\begin{lemma}
Let $M$ be a set of guards that see all points of a set of segments $S$.
Than $M$ sees all points of any subset of $S$.
\label{L3}
\end{lemma}
Follows from the observation that the removal of a segment $s \in S$ does not affect the visibility 
of the points of $S \setminus s$. 
Note that this does not always apply if an edge of $\bar G_S$ is removed (see Figure~\ref{f8}). 

\smallskip

With this preparation, we are ready to describe our algorithm and evaluate its complexity.

\subsection*{Algorithm for Guarding a Set of Segments with a Tree Structure} 

The input to the problem is a set of segments $S$ and its output is a minimal set $M$ of vertices guarding $S$.
Using the procedures (A), (B), (C), and (D), we can compute all necessary sets $V = I \cup J$, $S_u$, $E_s$, $F_u$, and $E_u$
for all vertices $u \in V$ and segments $s \in S$, as well as all appropriate vertices for $T$.
The algorithm consists of the following steps.

\smallskip

\noindent {\bf Guarding Plane Tree Algorithm}
\begin{enumerate}
\item[1.]
Choose an appropriate vertex $u$. 
If the graph has more than one connected component, the appropriate vertex can belong to an arbitrary component.
\item[2.]
Find its parent $v$ and place a guard in it (that is, store $v$ in a list $M$ of guards). 
\item[3.]
Remove $F_u$ from $E$. ({\em Note:} As commented above, the last action will disconnect $T$ 
and turn it to a forest with $|F_u| + 1$ components.)

If $E = \emptyset$, then stop and report $M$ as a solution to the problem. Otherwise go to Step 4.
\item[4.] 
Update the obtained graph as follows:

For every segment $s \in S_u$ and for every vertex $v$ of an edge in $E_s$, check if $|S_v| = 2$.
If this is the case, take the other segment $r$ through $v$. 
Using the list $E_r$, identify the two edges $(p,v)$, $(v,q)$ 
incident to $v$ and merge them into one edge $(p,q)$.
Go to Step 1.  
\end{enumerate}

\begin{theorem}
Let $S$ be a set of segments whose graph is a tree $T$ with $p$ vertices. 
The Guarding Plane Tree algorithm solves correctly the GSS problem with $O(p \log p)$ operations.
\label{th2}
\end{theorem}
{\bf Proof} \ {\em 1. Correctness} \ 
We prove  by strong mathematical induction on the number of edges of the graph.
The statement is obvious for a tree with one or two edges. 
Assume that it is true for any tree with $i$ edges, $3 \leq i \leq k$.
We will show that then it holds for a tree $T$ with $k+1$ edges.

Let $u$ be an appropriate leave of $T$ (which exists by Lemma~\ref{L1}),
and let $v$ be its parent.  
Following the Guarding Plane Tree algorithm, remove $F_v$ from $E$. 
Denote the obtained graph by $T'$.   
The latter is a forest with less than $k$ edges to which the inductional hypothesis applies. 
Therefore, proceeding with the algorithm on $T'$, we will obtain a minimal set 
$\Gamma'= \{ v_1,v_2,\dots,v_p \}$ of guards for $T'$. 

We have that:
\begin{itemize}
\item[-]
$v$ sees $F_v$ (and $S_v$, respectively);  
\item[-]
The vertices $v_1,v_2,\dots,v_p$ see $T'$ and possibly part of $F_v$ and $v$ itself, but do not see 
$u$ (as $u$ is appropriate) and the edge $(u,v)$.
\end{itemize}
Thus, the set of vertices $\Gamma = \{ v_1,v_2,\dots,v_p,v \}$ guards $T$.

What remains to verify is that $\Gamma$ is a minimal set of guards for $T$. 
Assume the opposite, i.e., that there is a set of guards for $T$ with less than $p+1$ elements.
Since $T'$ is obtained from $T$ by removal of the segments $S_v$,  the minimal number of  guards for $T$ cannot be less
than $p$. 
Assume then that there are vertices $u_1,u_2,\dots,u_{p-1},u_p$ that guard $T$. 
Since $v$ is the only vertex that can see $u$, one of the above must be $v$, e.g., $u_p=v$. 
Since $v$ can see only the edges of $F_u$ (segments $S_u$, respectively), then the rest of the tree 
must be guarded by the other $p-1$ vertices $u_1,u_2,\dots,u_{p-1}$. 
This contradicts the minimality of $\Gamma'= \{ v_1,v_2,\dots,v_p \}$. 

\smallskip

{\em 2. Running Time} \ 
By the algorithm description it is clear that, once we have computed 
$V = I \cup J$, $S_u$, $E_s$, $F_u$, $E_u$, and the set of appropriate leaves,
the five steps of the algorithm can be performed in $O(n)$ time. 
Thus the overall algorithm time complexity amounts to $O(p \log p)$. 
\qed

\section{Concluding Remarks}
\label{concl}
In this paper we considered the problem of finding a minimal number of guards 
that can guard a set of segments in the plane.
We proved that the problem is strongly NP-complete even for sets of segments with a cubic graph structure.
We also designed a polynomial algorithm for the case when the graph associated to the set of segments is a tree.

Work in progress is aimed at investigating (both theoretically and experimentally)
the approximability of the considered problem.
It is well-known that a minimal set cover can be found in polynomial time within an $O(\log n)$ factor, 
which is the best possible by order (unless the problems in NP admit quasi-polynomial time solutions). 
A minimal vertex cover can efficiently be computed within a constant factor.
What an approximation is possible GSS?

\section*{Acknowledgements}

This work was supported in part by NSF grant No 0802964.


\end{document}